\documentclass[runningheads]{llncs}

\usepackage[mobile]{eccv}
\usepackage{eccvabbrv}
\usepackage{graphicx}
\usepackage{booktabs}
\usepackage{algorithm}
\usepackage{algorithmic}
\usepackage{booktabs} 
\usepackage{amsmath}
\usepackage{multicol}
\usepackage{multirow}
\usepackage{xcolor}
\usepackage{amsfonts}
\usepackage{utfsym}
\usepackage[accsupp]{axessibility}

\usepackage[pagebackref,breaklinks,colorlinks,citecolor=accvblue]{hyperref}
\usepackage{orcidlink}

\begin{document}

\title{Mamba-based Light Field Super-Resolution \\ with Efficient Subspace Scanning} 

\titlerunning{MLFSR}

\author{Ruisheng Gao \and
Zeyu Xiao \and
Zhiwei Xiong
}

\authorrunning{Gao et al.}
\institute{University of Science and Technology of China}

\maketitle

\begin{abstract}
Transformer-based methods have demonstrated impressive
performance in 4D light field (LF) super-resolution by effectively modeling
long-range spatial-angular correlations, but their quadratic complexity
hinders the efficient processing of high resolution
4D inputs, resulting in slow inference speed and high memory
cost. As a compromise, most prior work adopts a patch-based strategy,
which fails to leverage the full information from the entire input LFs.
The recently proposed selective state-space model, Mamba, has gained
popularity for its efficient long-range sequence modeling. In this paper, we propose a Mamba-based Light Field
Super-Resolution method, named MLFSR, by designing an efficient subspace
scanning strategy. Specifically, we tokenize 4D LFs into subspace sequences
and conduct bi-directional scanning on each subspace. Based on our scanning
strategy, we then design the Mamba-based Global Interaction (MGI)
module to capture global information and the local Spatial-
Angular Modulator (SAM) to complement local details. Additionally, we introduce a
Transformer-to-Mamba (T2M) loss to further enhance overall performance. Extensive
experiments on public benchmarks demonstrate that MLFSR
surpasses CNN-based models and rivals Transformer-based methods in
performance while maintaining higher efficiency. With quicker inference speed
and reduced memory demand, MLFSR facilitates
full-image processing of high-resolution 4D LFs with enhanced performance.
  
\end{abstract}

\section{Introduction}
Light field (LF) imaging technique enables the capture of both intensity and direction of light rays, therefore finding wide applications in virtual reality and computational photography. 
However, the design of commercialized LF cameras (\textit{e.g.}, Lytro) is mainly based on the micro-lens array. This design divides the main lens into sub-apertures, introducing a trade-off between the spatial and angular resolution of LFs. 
Therefore, LF Super-Resolution (LFSR) has become an important research topic to address these challenges.

The spatial-angular redundancy in the 4D LF structure is vital for LFSR methods to exploit, compensating for missing high-frequency details in each view.
Deep learning techniques, including CNNs and Transformer-based architectures, have been widely adopted and proven effective for modeling 4D correlations within LFs.
Directly processing entire 4D LF representations, such as Sub-Aperture Images (SAI) or Macro-Pixel Images (MacPI), poses optimization challenges during the training stage~\cite{meng2019high}.
Consequently, prior methods extract features from 2D slices, including spatial slices, angular slices, and Epipolar Plane Images (EPIs)~\cite{wang2020spatial,wang2022disentangling,cheng2022spatial,van2023light,wang2022detail,liang2022light,liang2023learning,cong2023exploiting}. 
However, CNN-based methods, with their limited receptive fields, struggle to capture long-range spatial-angular correspondences, leading to inferior performance. Although Transformer-based methods effectively model non-local information, their attention mechanisms incur high memory costs and slow inference time.

Another critical aspect is the inference scheme. 
The drawbacks mentioned earlier have led prior works to adopt a patch-inference scheme, where input low-resolution (LR) LFs are divided into patches for individual processing, and the outputs are merged to form the final super-resolved result. 
This approach has two main issues: (1) For full-resolution LFs, the number of patches increases, requiring multiple forward passes through the network. 
(2) Each LF patch carries limited information, preventing the network from fully leveraging the entire input LF~\cite{chu2022improving}. 
Therefore, there is an urgent need for efficient architectures that consume less memory and offer faster inference speeds to process full-resolution input LFs, thereby fully utilizing the whole input 4D LF structure.

The currently trending state space models (SSMs)\cite{gu2021efficiently,smith2022simplified,gu2023mamba} have proven effective in modeling long-range information with theoretically linear complexity. 
Among these, Mamba\cite{gu2023mamba} introduces a selective mechanism for initializing structural parameters, enabling context-aware sequence modeling and boosting performance. Inspired by Mamba's efficient long-range modeling capability, we explore its potential in handling the complex 4D structure of LFs.

\begin{figure}[t!]
    \centering
    \includegraphics[width=0.88\linewidth]
    {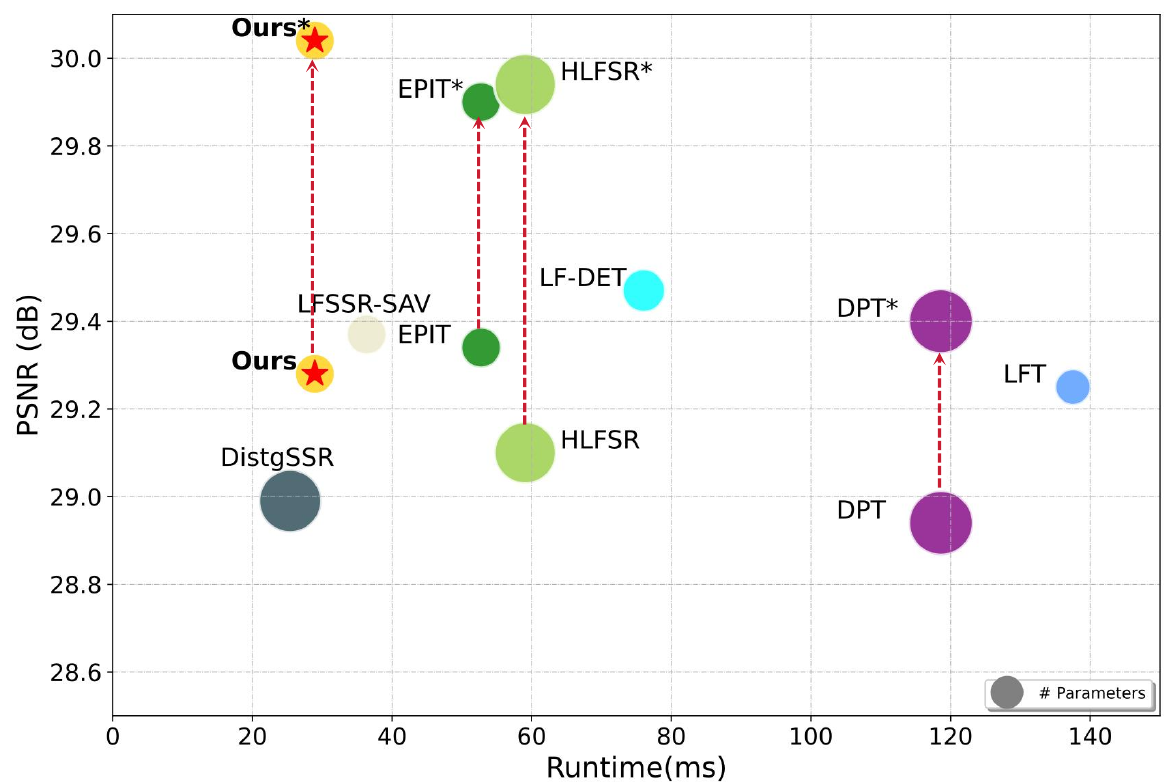}
    \caption{Runtime (ms) and PSNR (dB) comparison. The runtime is calculated on an input LF of size 5$\times$5$\times$32$\times$32, * denotes full-resolution inference.
    MLFSR outperforms state-of-the-art CNN-based methods and has competitive performance compared to Transformer-based methods with less runtime and parameters. 
    Full-resolution inference gives a further performance boost on available methods. 
    Comparisons are performed for 4$\times$ SR on the EPFL dataset. }
    \label{Fig:teaser}
\end{figure}

In this paper, we attempt to leverage the efficient merits of Mamba to overcome the limitations of CNNs and Transformer-based methods.
Specifically, we consider the inherent redundancy in the 4D LF structure and propose an efficient bi-directional subspace scanning scheme to effectively capture long-term spatial-angular correspondences.
Based on the above scanning scheme, we propose a Mamba-based Global Interaction (MGI) module for the spatial-angular representation (SA-Mamba) and EPI representation (EPI-Mamba).
Additionally, we propose a Spatial-Angular Modulator (SAM) to preserve the local structure of processed features.
Combining the above designs, we devise a Mamba-based LFSR network termed MLFSR, which comprises interleaved MGI and SAM to efficiently model the 4D LF structure.
In practice, there's still some performance gap between MLFSR and Transformer-based methods due to the lossy state compression in SSMs.
We therefore propose a Transformer-to-Mamba (T2M) distillation loss to align the non-local modelling ability between the attention mechanism and SSMs to improve the overall performance.
As demonstrated in Figure~\ref{Fig:teaser}, experimental results show that MLFSR outperforms CNN-based methods and benefits from less runtime and memory cost compared to Transformer-based counterparts while achieving competitive performance.

The contributions of this paper are summarized below.
\begin{itemize}
\item We consider the spatial-angular redundancy that existed in 4D LFs and propose a bi-directional subspace scanning scheme to efficiently enable global information utilization.  
\item Based on the scanning scheme, we propose an MGI module and a local SAM to model complex 4D correlations efficiently, forming the MLFSR.
We then introduce a T2M distillation loss to further boost the overall performance.
\item Our proposed MLFSR outperforms CNN-based methods while comparable with Transformer-based counterparts while enjoying lower memory cost and less inference time, which enables full-resolution inference for further performance improvement.
\end{itemize}

\section{Related Work}
\subsection{Light Field Super-Resolution}
Representative traditional LFSR methods attempt to utilize depth information under optimization frameworks~\cite{bishop2009light,bishop2011light} or learning from dictionaries~\cite{cho2013modeling} to ensure angular consistency.
With the advancement of deep learning techniques, CNN-based methods have gradually become the mainstream choice.
The pioneer work LFCNN~\cite{yoon2015learning} uses stacked convolutional layers to process vertical, horizontal, and surrounding sub-aperture images.
Following methods adopt 4D convolutions on SAIs to extract spatial-angular features~\cite{meng2019high}. Yeung et al.~\cite{yeung2018light} further ease the computational burden by utilizing spatial-angular separable convolutions. 
To well-exploit the entangled 4D information, later works disentangle the LF into 2D subspaces and design convolution variants tailored for each subspace~\cite{wang2020spatial,wang2022disentangling}.
HLFSR~\cite{van2023light} further proposes to leverage multi-orientation epipolar information to boost the performance.
However, due to the limited receptive field of CNN, non-local spatial-angular information cannot be well incorporated.
Liu \textit{et. al}~\cite{liu2023light} design a selective matching mechanism to explicitly extract 4D information in LFs with larger receptive fields.
Recently, Transformer-based methods have emerged and achieved superior performance.
By tokenizing 2D slices into sequences, the attention mechanism is adopted in spatial-angular~\cite{liang2022light} or EPI~\cite{liang2023learning} subspaces to learn non-local spatial-angular correlations.
Instead of network design, Xiao \textit{et. al}~\cite{xiao2023cutmib} improves the SR performance from a data augmentation perspective.
Different from prior works, we introduce the newly proposed Mamba to overcome the drawback of limited receptive field in CNN-based methods and the inferior efficiency of Transformer-based methods. 

\subsection{State Space Models}
Originating from control theory, SSMs have garnered increasing attention due to their efficacy in long-term language modeling~\cite{gu2021efficiently,smith2022simplified,gu2023mamba}. Unlike self-attention-based transformers, most SSMs capture long-range token interactions through linear recurrent processes, entailing $\mathcal{O}(N)$ complexity theoretically. 
Mamba~\cite{gu2023mamba} improves the expressiveness of SSMs by introducing a selective mechanism, with its structural parameters adaptively learned from inputs. 
Motivated by its effectiveness, researchers have endeavored to adapt Mamba to the domain of computer vision. 
Vim~\cite{zhu2024vision} and VMamba~\cite{liu2024vmamba} take pioneering efforts employing multidirectional scanning to overcome the causal nature of vanilla Mamba, inspiring subsequent work in various vision domains~\cite{guo2024mambair,xing2024segmamba,peng2024fusionmamba,huang2024localmamba,li2024videomamba}. 
In this paper, we investigate the applicability of Mamba to LFs, to unlock its potential in efficient LF processing.

\section{Method}
\subsection{Preliminaries}
SSMs are inspired by the continuous control systems, which map input sequence $x(t) \in \mathbb{R}^{d}$ to output sequence $y(t) \in \mathbb{R}^{d}$ through hidden state $h(t) \in \mathbb{R}^{n}$.
Without nonlinear transformation, SSMs set linear mappings $\boldsymbol{\mathrm{A}}$, $\boldsymbol{\mathrm{B}}$ and $\boldsymbol{\mathrm{C}}$ as transition parameters to update the hidden state over time and obtain the output.
The mapping process is as follows,

\begin{equation}
\label{Eq:ssmconti}
\begin{aligned}
  h'(t)& = \boldsymbol{\mathrm{A}}h(t) + \boldsymbol{\mathrm{B}}x(t),  \\
  y(t) & = \boldsymbol{\mathrm{C}}h(t).
\end{aligned}
\end{equation}

To enable deep network training, recent SSM such as the Structure State Space Sequence model (S4)~\cite{gu2021efficiently} and Mamba~\cite{gu2023mamba} discretize the above process to a discretized version through an extra timescale parameter $\boldsymbol{\mathrm{\Delta}}$.
Then, continuous parameters $\boldsymbol{\mathrm{A}}$ and $\boldsymbol{\mathrm{B}}$ are transformed into the discretized ones $\overline{\boldsymbol{\mathrm {A}}} $ and $\overline{\boldsymbol{\mathrm {B}}} $ via the commonly used zero-order hold (ZOH) method,

\begin{equation}
\label{Eq:zoh}
\begin{aligned}
  \overline{\boldsymbol{\mathrm {A}}}& = \mathrm {exp}(\boldsymbol{\mathrm{\Delta}}\boldsymbol{\mathrm{A}}) ,  \\
  \overline{\boldsymbol{\mathrm {B}}} & = (\boldsymbol{\mathrm{\Delta}}\boldsymbol{\mathrm{A}})^{-1}  (\mathrm {exp}(\boldsymbol{\mathrm{\Delta}}\boldsymbol{\mathrm{A}})-\boldsymbol{\mathrm{I}}) \boldsymbol{\mathrm{\Delta}}\boldsymbol{\mathrm{B}},
\end{aligned}
\end{equation}
where $\boldsymbol{\mathrm{I}}$ denotes the identity matrix and $(\cdot)^{-1}$ is the matrix inversion operation.
Therefore, we now derive the discretized version of E.q.~\ref{Eq:ssmconti},
\begin{equation}
\label{Eq:ssmdiscre}
\begin{aligned}
  h_{t}& = \overline{\boldsymbol{\mathrm {A}}}h_{t-1} + \overline{\boldsymbol{\mathrm {B}}}x_{t},  \\
  y_{t} & = \boldsymbol{\mathrm {C}}h_{t}.
\end{aligned}
\end{equation}

The above recurrent form can be deviated into a convolutional form with global kernel $\boldsymbol{\mathrm {K}}$, which enables fast parallel training,
\begin{equation}
\label{Eq:ssmconv}
\begin{aligned}
\overline{\boldsymbol{\mathrm {K}}} &  = (\boldsymbol{\mathrm {C}}\overline{\boldsymbol{\mathrm {B}}},\boldsymbol{\mathrm {C}}\overline{\boldsymbol{\mathrm {AB}}},...,\boldsymbol{\mathrm {C}}\overline{\boldsymbol{\mathrm {A}}}^{L-1}\boldsymbol{\mathrm {B}}), \\
\boldsymbol{\mathrm {y}} &  = \boldsymbol{\mathrm {x}} \ast \boldsymbol{\overline{\mathrm {K}}},
\end{aligned}
\end{equation}
where $L$ denote the sequence length and $\ast$ represent the convolution operator.

Different from S4, Mamba introduces a selective mechanism on $\overline{\boldsymbol{\mathrm {B}}}$ and $\boldsymbol{\mathrm {C}}$ to enable context-aware information filtering.
Although it breaks the available convolutional form by introducing nonlinearity, a parallel scan algorithm~\cite{gu2023mamba} is proposed to speed up the process in E.q.~\ref{Eq:ssmdiscre} for faster training.

\begin{figure}[t!]
    \centering
    \includegraphics[width=\linewidth]
    {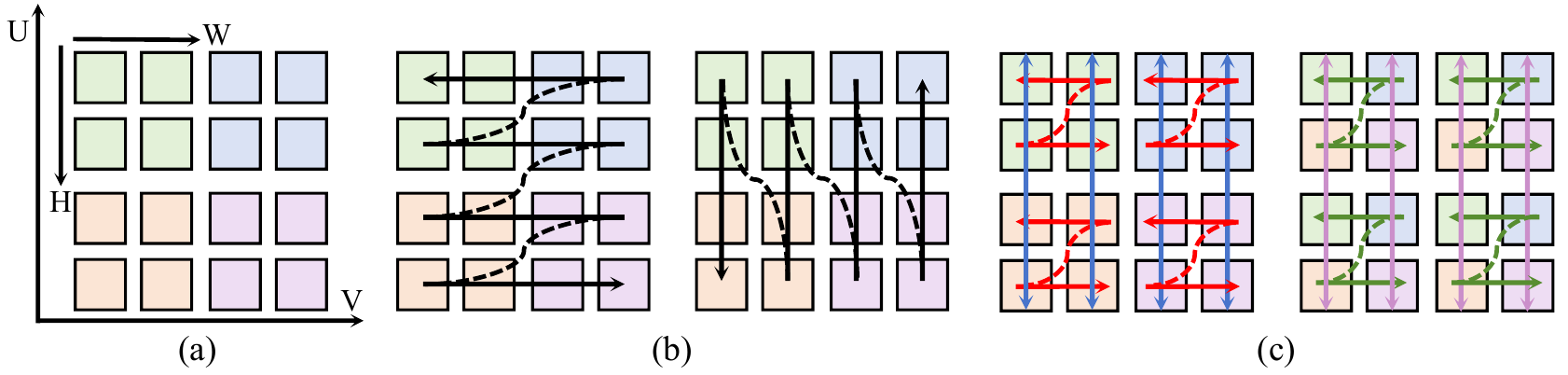}\
    \vspace{-3mm}
    \caption{A toy example ($U=V=2,H=W=2$) on LF tokenization and scanning directions. (a) Sub-aperture images. (b) Whole 4D sequence with quad-directional scanning (SAI for example). (c) Subspace sequences with bi-directional scanning. Bi-directional spatial (red arrows) and angular (green arrows) scanning are complemented by bi-directional EPI (blue and purple arrows) scanning.}
    \label{Fig:scanning}
\end{figure}

\subsection{Mamba-based Light Field Super-Resolution}
\begin{figure}[t!]
    \centering
    \includegraphics[width=1\linewidth]
    {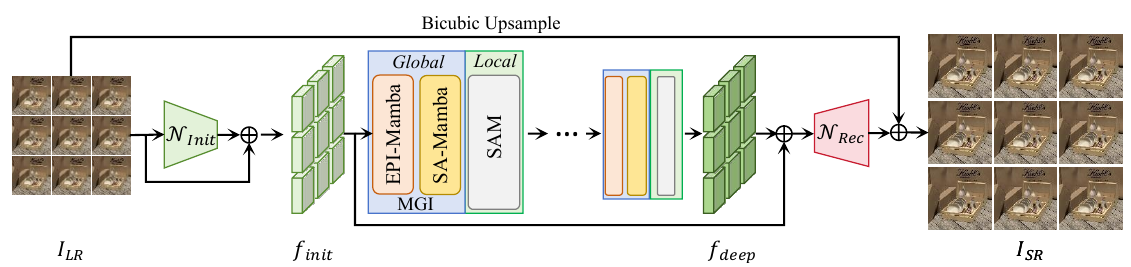}
    \vspace{-4mm}
    \caption{Overview of MLFSR. Initial features $f_{init}$ are first extracted by encoder $\mathcal{N}_{Init}$, followed by alternate MGI modules and SAM to extract deep features $f_{deep}$. Finally, we obtain super-resolved results $I_{SR}$ through the reconstruction module $\mathcal{N}_{Rec}$. }
    \label{Fig:pipeline}
    \vspace{-1mm}
    
\end{figure}

\noindent\textbf{Overview.} 
Taking an input LR LF $I_{LR}\in \mathbb{R}^{U\times V\times H\times W}$, we aim to increase its spatial resolution by scale $s$, resulting in a super-resolved output $I_{SR}\in \mathbb{R}^{U\times V\times sH\times sW} $, where $U,V$ denotes the angular resolution and $H, W$ denotes the spatial resolution. 
As shown in Fig.~\ref{Fig:pipeline}, we first extract shallow features $f_{init}$ from LR LF input $I_{LR}$ using a convolutional-based encoder $\mathcal{N}_{Init}$.
Then the shallow features are fed to alternate global-local 4D correspondence learning modules including Mamba-based Global Interaction
(MGI) module and Spatial-Angular Modulator (SAM) to extract informative deep features $f_{deep}$.
Finally, $f_{deep}$ is upsampled by the reconstruction module $\mathcal{N}_{Rec}$ to obtain the upsampled LF $I_{SR}$.

\noindent\textbf{Efficient Subspace Scanning.} 
As a high-dimensional vision modality, 4D LF has various representation forms such as SAIs, MacPIs, and subspace images.
Therefore, it's vital to select an appropriate way for effective and efficient scanning of specific representations.
A straightforward way is to directly conduct 2D quad-directional scanning on the whole 4D representation (SAI or MacPI), as shown in Fig.\ref{Fig:scanning}(b).
In this way, the tokenized sequence has $U\times V\times H\times W$ length and the efficient merit of Mamba on long sequence inputs can be fully exploited.
Considering the causal nature of the original Mamba~\cite{gu2023mamba}, quad-directional scanning~\cite{liu2024vmamba} thus is required to fully benefit each token from each view.
However, as it's important in LFSR to well leverage the intra-view and inter-view correspondences, the mixed intra-inter view information cannot be well distinguished for each token.
On the other hand, the compressed hidden state struggles to preserve prior spatial-angular information when the sequence gets longer.
The trade-off between the long-range efficiency of Mamba and the limited state memory provides challenges to well adopting Mamba for LF inputs.

To overcome the above dilemma, we attempt to utilize the inherent structure redundancy that exists in LFs.
Specifically, we notice the role of EPI representation, which provides rich information across the spatial and angular dimensions.
When conducting bi-directional scanning on the EPI-H (or EPI-W) subspace, the vertical (or horizontal) of spatial and angular information can be incorporated, as can be seen in Fig.\ref{Fig:scanning}(c).
The quad-directional spatial (or angular) scanning can therefore be decomposed into two bi-directional scanning on spatial (or angular dimensions, marked by red and green arrows) and EPI dimensions (marked by blue and purple arrows), respectively. 
In this way, by lowering the scanning length, the long-term memory issues can be well handled without sacrificing the complete 4D global information.

\begin{figure}[t!]
    \centering
    \includegraphics[width=0.88\linewidth]
    {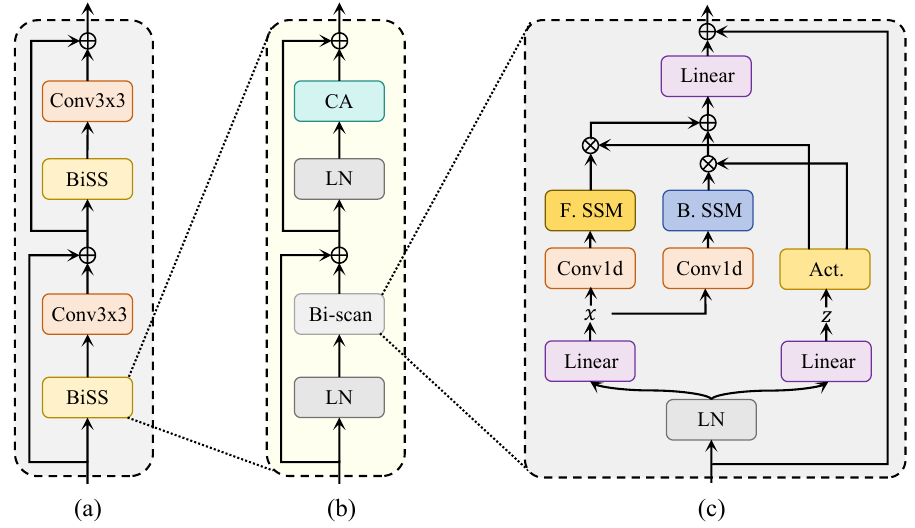}
    \vspace{-1mm}
    \caption{(a) The detailed structure of SA-Mamba/EPI-Mamba. Each SA-Mamba/EPI-Mamba includes two Bidirectional Subspace Scanning (BiSS) blocks. We omit the reshape operation for simplicity. (b) The BiSS block follows a typical Transformer-style design, including bidirectional scanning for token interaction and channel attention for channel mixing. (c) The details of bidirectional scanning. }
    \label{Fig:MGI}
    \vspace{-1mm}
    
\end{figure}

\noindent\textbf{Mamba-based Global Interaction (MGI) Module.} 
Based on the above bi-directional subspace scanning, we design a Mamba-based Global Interaction (MGI) module which consists of cascaded EPI-Mamba and SA-Mamba for effective non-local spatial-angular correspondence learning, as shown in Fig.~\ref{Fig:pipeline}.
We take EPI-Mamba for example to elaborate its details since the structures of EPI-Mamba and SA-Mamba are the same except for the input form.

The detailed structures are illustrated in Fig.~\ref{Fig:MGI}(a).
Given an input feature $f^{i-1}_{l}$ from the output of i-1-th SAM, we first flatten it into EPI-H token sequences $T^{i}_{h}\in \mathbb{R}^{BVW\times UH\times C}$, where $C$ is the channel dimension and $B$ is the batch size.
Then, two successive Bidirectional Subspace Scanning (BiSS) blocks and convolution layers are set to enable non-local token interaction on the EPI-H and EPI-W subspace, which can be formulated as follows,
\begin{equation}
	\begin{aligned}
			\hat{T^{i}_{h}} & =\mathrm {Conv}(\mathrm{BiSS}(T^{i}_{h})) + T^{i}_{h}, \\
					\hat{T^{i}_{w}} & =\mathrm {Conv}(\mathrm{BiSS}(T^{i}_{w})) + T^{i}_{w},
	\end{aligned}
\end{equation}

where the EPI-W token sequences $T^{i}_{w}\in \mathbb{R}^{BUH\times VW\times C}$ is reshaped from $\hat{T^{i}_{h}}$ and $\hat{T^{i}_{h}}$ and $\hat{T^{i}_{w}}$ are enhanced features.
$\hat{T^{i}_{w}}$ is then reshaped to $f^{i}_{g}$ as the output of the i-th MGI module.
The design of the BiSS block follows a typical Transformer-style structure, which is shown in Fig.~\ref{Fig:MGI}(b).
Input token sequences are first normalized by a LayerNorm~\cite{ba2016layer} and fed into a Bidirectional Scanning (Bi-scan) block~\cite{zhu2024vision} (see Fig.~\ref{Fig:MGI}(c)).
Then, another LayerNorm and a Channel Attention (CA) layer~\cite{zhang2018image} is used for channel mixing.
We also add shortcut connections~\cite{he2016deep} to ease the optimization process.

As for the SA-Mamba, we can simply replace EPI token sequences $T^{i}_{h}$ and $T^{i}_{w}$ with spatial and angular sequences $T^{i}_{s} \in \mathbb{R}^{BUV\times HW\times C}$ and $T^{i}_{a}\in \mathbb{R}^{BHW\times UV\times C}$.
In practice, we share the weights of two BiSS and convolutional layers in EPI-Mamba to share the intrinsic information along the epipolar line of LFs~\cite{liang2023learning}, which also leads to more efficient network architecture.

By combining the EPI-Mamba and SA-Mamba, complementary information between two directions of spatial-angular information and EPI information can be well incorporated, achieving efficient global 4D correspondence learning.
The experimental results in Section~\ref{Sec:ablation} validate the effectiveness of each component.

\begin{figure}[t!]
    \centering
    \includegraphics[width=0.7\linewidth]
    {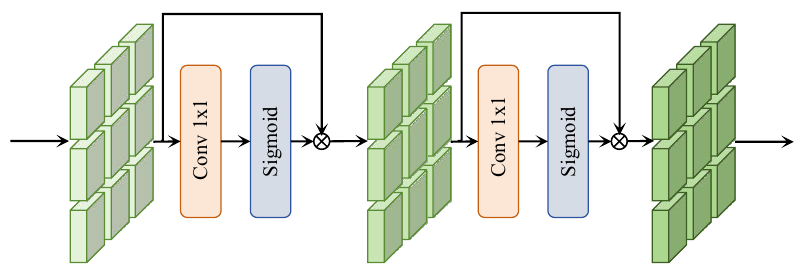}
    \caption{The detailed structure of Spatial-Angular Modulator (SAM).}
    \label{Fig:sam}
    \vspace{-2mm}
    
\end{figure}

\noindent\textbf{Spatial-Angular Modulator (SAM).} 
Although MGI enables the global receptive field to enable the network to utilize non-local spatial-angular information, the importance of locality in SR~\cite{chen2023dual} is overlooked.
To compensate for local spatial-angular information, we design a convolution-based Spatial-Angular Modulator (SAM) to learn input-adaptive weights for modulating the output of MGI.
Inspired by pixel attention~\cite{zhao2020efficient}, we adopt a lightweight design of SAM with only two convolutional layers with kernel size 1, as shown in Fig~\ref{Fig:sam}.

The output feature $f^{i}_{g}\in \mathbb{R}^{B\times U\times V\times H\times W\times C}$ from last MGI module is first reshaped to the spatial subspace and go through a single convolutional layer, followed by a sigmoid operation to generate the spatial attention map $\mathrm{Attn}_{s}$.
Then, $f^{i}_{g}$ is modulated using $\mathrm{Attn}_{s}$ by element-wise multiplication,
\begin{equation}
\label{Eq:sam1}
\begin{aligned}
    \mathrm{Attn}_{s} & = \mathrm {Sigmoid}(\mathrm {Conv}(f^{i}_{g})),\\
    \tilde{f^{i}_{g}} & = f^{i}_{g} \cdot \mathrm{Attn}_{s} + f^{i}_{g},
\end{aligned}
\end{equation}
where $\tilde{f^{i}_{g}}$ represents spatially-modulated features.
A similar process is conducted on the angular subspace to obtain the corresponding angular attention map $\mathrm{Attn}_{a}$ and we can obtain the output $f^{i}_{l}$,
\begin{equation}
\label{Eq:sam2}
\begin{aligned}
    f^{i}_{l} & = \tilde{f^{i}_{g}} \cdot \mathrm{Attn}_{a} + \tilde{f^{i}_{g}}.
\end{aligned}
\end{equation}

\subsection{Learning Objective}
We use $L_1$ distance to measure the reconstruction loss between the final output and the ground-truth LF,

\begin{equation}
\label{Eq:loss1}
\begin{aligned}
\mathcal{L} _{rec} = \left \| I_{SR}-I_{HR} \right \| _{1} .
\end{aligned}
\end{equation}
However, only using the reconstruction loss results in sub-optimal performance compared to Transformer-based methods according to the experiment results.
Considering the difference between the attention mechanism and SSMs, we can find that attention-based methods explicitly record relations between tokens into the attention map, while SSMs implicitly compress non-local information in hidden memory.
We therefore propose a Transformer-to-Mamba (T2M) distillation loss to align the deep features of the Transformer and Mamba, which is formulated as follows,

\begin{equation}
\label{Eq:loss1}
\begin{aligned}
\mathcal{L} _{dist} = \left \| f_{deep}-f^{Trans}_{deep} \right \| _{1} .
\end{aligned}
\end{equation}

In this work, we utilize EPIT~\cite{liang2023learning} as the teacher network. 
The total loss function is the combination of the above two loss functions, adjusted by $\lambda$,

\begin{equation}
\label{Eq:loss1}
\begin{aligned}
\mathcal{L} _{total} = \mathcal{L} _{rec} + \lambda\mathcal{L} _{dist}.
\end{aligned}
\end{equation}

\begin{table*}[t!]
\centering
\caption{Quantitative results (PSNR / SSIM) on the five public benchmarks of LF super-resolution with the input angular resolution of 5$\times$5. We report the number of parameters (Params.) and runtime for efficiency comparison. Note that the runtime is calculated on an input LF of size 5$\times$5$\times$32$\times$32. Method marked with * denotes full-resolution inference results. We mark the best, second best results in \textbf{bold} and with \underline{underline}, respectively.}
\label{Tab:SR}
\scalebox{0.75}{
\begin{tabular}{@{}lc|cc|ccccc}
\toprule
Method & Scale & Params.(M) & Time (ms) & \multicolumn{1}{c}{HCI-new} & \multicolumn{1}{c}{HCI-old} & \multicolumn{1}{c}{EPFL} & \multicolumn{1}{c}{INRIA} & \multicolumn{1}{c}{STFGantry} \\ \midrule

Bicubic & 2$\times$ & - & - & 31.89/.9356  & 37.69/.9785 & 29.74/.9376  & 31.33/.9577  & 31.06/.9498  \\ \midrule
VDSR & 2$\times$ & 0.66 & 29.50 &   34.37/.9561   &  40.61/.9867   & 32.50/.9598  & 34.43/,9741  & 35.54/.9789  \\
EDSR & 2$\times$ & 38.62  &  137.25 &   34.83/.9592   &  41.01/.9874   & 33.09/.9629  &  34.97/.9764  & 36.29/.9818  \\
RCAN & 2$\times$ & 15.31 & 1586.50 &  34.98/.9603  &   41.05/.9875   & 33.16/.9634  & 35.01/.9769  & 36.33/.9831 \\
resLF & 2$\times$ & 7.98 & 101.89 &  36.69/.9739   &   43.42/.9932   & 33.62/.9706  &  35.39/.9804  & 38.36/.9904 \\
LFSSR & 2$\times$ & 0.89 & 3.98 & 36.81/.9749   &  43.81/.9938   & 33.68/.9744  &  35.28/.9832  &  37.95/.9898  \\
LF-InterNet & 2$\times$ & 5.04 & 18.08 &   37.28/.9763  &   44.45/.9946   & 34.14/.9760  & 35.80/.9843  & 38.72/.9909  \\
DistgSSR & 2$\times$ & 3.53 & 24.76 & 37.96/.9796  &   44.94/\underline{.9949}   & 34.81/.9787  & 36.59/.9859  & 40.40/.9942  \\
LFSAV & 2$\times$ & 1.22 & 10.57 & 37.43/.9776  &   44.22/.9942   & 34.62/.9772  & 36.36/.9849  &  38.69/.9914  \\
HLFSR & 2$\times$ & 3.45 & 54.29 &  38.03/.9798  &  44.86/\underline{.9949}  &  34.71/.9779  & 36.59/.9856  & 40.49/.9943 \\ 
\midrule

LFT & 2$\times$ & 1.11 & 132.37 & 37.84/.9791   &  44.52/.9945   & 34.80/.9781  & 36.59/.9855  & 40.51/.9941   \\
DPT & 2$\times$ & 3.73 & 117.97 &   37.35/.9771   &  44.31/.9943   & 34.48/.9758  & 36.40/.9843  & 39.52/.9926   \\
LF-DET & 2$\times$ & 1.59 & 73.97 & \textbf{38.31}/\underline{.9807}  &   44.99/\textbf{.9950}   & \underline{35.26}/.9797  & \underline{36.95}/\underline{.9864}  & \underline{41.76}/\underline{.9955}  \\ 
EPIT & 2$\times$ & 1.42 & 51.58  & \underline{38.23}/\textbf{.9810}  &   \textbf{45.08}/\underline{.9949}   & 34.83/.9775  & 36.67/.9853  & \textbf{42.17}/\textbf{.9957} \\ \midrule

MLFSR (Ours) & 2$\times$ & 1.36 & 27.81  & 38.14/.9803 & 44.90/\textbf{.9950} & 35.22/\textbf{.9801}  & 36.92/\textbf{.9865}  &  40.98/.9949 \\ 
MLFSR* (Ours) & 2$\times$ & 1.36 & 27.81  &  38.15/.9802  & 44.97/\textbf{.9950} & \textbf{36.01}/\underline{.9798}  & \textbf{38.65}/\underline{.9864}  & 41.03/.9947  \\ \midrule \midrule

Bicubic & 4$\times$ & - & - & 27.61/.8517  & 32.42/.9344 & 25.14/.8324  & 26.82/.8867  &  25.93/.8452  \\ \midrule
VDSR & 4$\times$ & 0.66 & 35.75 &   29.31/.8823   &  34.81/.9515   & 27.25/.8777  & 29.19/.9204  & 28.51/.9009  \\
EDSR & 4$\times$ & 38.89  & 147.00 &   29.60/.8869   &  35.18/.9536   & 27.84/.8854  & 29.66/.9257  & 28.70/.9072  \\
RCAN & 4$\times$ & 15.36 & 1716.05 &   29.63/.8886   &  35.20/.9548   & 27.88/.8863  & 29.76/.9276  & 28.90/.9131  \\
resLF & 4$\times$ & 8.65 & 104.23 &   30.73/.9107   &  36.71/.9682   & 28.27/.9035  & 30.34/.9412  &  30.19/.9372  \\
LFSSR & 4$\times$ & 1.77 & 22.37 & 30.72/.9145   &  36.70/.9696   & 28.27/.9118  &  30.31/.9467  & 30.15/.9426  \\
LF-InterNet & 4$\times$ & 5.48 & 19.65 &   30.98/.9161   &  37.11/.9716   & 28.67/.9162  & 30.64/.9491  & 30.53/.9409  \\
DistgSSR & 4$\times$ & 3.58 & 25.40 & 31.38/.9217 & 37.56/.9732   & 28.99/.9195  & 30.99/.9519 & 31.65/.9535  \\
LFSAV & 4$\times$ & 1.54 & 36.36 & 31.45/.9217 & 37.50/.9721   & 29.37/\underline{.9223}  & 31.27/\underline{.9531} & 31.36/.9505  \\
HLFSR & 4$\times$ & 3.48 & 59.08 & 31.43/.9226 & 37.72/.9738 & 29.10/.9212 & 31.17/\underline{.9531}  & 31.41/.9524 \\ 
\midrule
LFT & 4$\times$ & 1.16 & 137.53 & 31.46/.9218 & 37.63/.9735   & 29.25/.9210  & 31.20/.9524 & 31.86/.9548  \\
DPT & 4$\times$ & 3.78 & 118.62 & 31.20/.9188 & 37.41/.9721   & 28.94/.9170  & 30.96/.9503 & 31.15/.9488  \\
LF-DET & 4$\times$ & 1.69 & 76.05 & \underline{31.56}/\underline{.9235} &  \underline{37.84}/.9744   & \underline{29.47}/\textbf{.9230}  & \underline{31.39}/\textbf{.9534}  & 32.14/\underline{.9573}  \\
EPIT & 4$\times$ & 1.47 & 52.75  & 31.51/.9231 & 37.68/.9737   & 29.34/.9197  & 31.37/.9526 & \underline{32.18}/.9571 \\  \midrule 
MLFSR (Ours)& 4$\times$ & 1.41 & 28.94 & \underline{31.56}/\underline{.9235} & 37.83/\underline{.9745}   & 29.28/.9218  & 31.24/\underline{.9531}  & 32.03/.9567  \\ 
MLFSR* (Ours) & 4$\times$ & 1.41 & 28.94 &  \textbf{31.59}/\textbf{.9237}  &   \textbf{37.92}/\textbf{.9747}   & \textbf{30.04}/.9221  & \textbf{32.21}/\textbf{.9534}  & \textbf{32.21}/\textbf{.9575}  \\

\bottomrule

\end{tabular}
}
\end{table*}

\section{Experiments}

\subsection{Experiment Settings}
\textbf{Datasets}. 
Aligning with prior works, we use the BasicLFSR benchmark~\cite{NTIRE2023LFSR} for the training and evaluation of all methods on 2$\times$ and 4$\times$ scale. 
The benchmark comprises five LF datasets (HCI-new~\cite{honauer2017dataset}, HCI-old~\cite{wanner2013datasets}, EPFL~\cite{rerabek2016new}, INRIA~\cite{le2018light} and STFGantry~\cite{StfGantry}) which include 144 training scenes and 23 test scenes with diverse contents and varying disparities.
We extract the central 5$\times$5 SAIs from each LF for training and testing.
\\

\noindent\textbf{Implementation Details}. 
We implement our method using Pytorch~\cite{paszke2019pytorch} framework with four NVIDIA 3090 GPUs.
The result of baseline methods is obtained from open-source codes.
The number of MGI and SAM used in MLFSR is set to 3 and the state dimension is 8.
The hyperparamter in $\mathcal{L} _{total}$ is set to 0.1.
The MLFSR is trained using the Adam optimizer with $\beta_{1}=$ 0.9 and $\beta_{2}=$ 0.999 with batch size 4 on each card.
The initial learning rate is set to 4e-4 and decreases by a factor of 0.5 for every 15 epochs.
Data augmentation including random flipping and rotation is applied in the training stage.
The network is trained for 90 epochs with L1 loss only and finetuned for another 30 epochs with a combination of L1 loss and T2M distillation loss.
We use widely adopted Peak Signal-to-Noise Ratio (PSNR) and Structure Similarity (SSIM) to evaluate the fidelity of the super-resolved results.

\begin{table}[!t]
\caption{Efficiency comparison against state-of-the-art Transformer-based methods on $2\times$ scale. We take runtime (ms) and peak memory use (GB) on different input spatial resolutions to evaluate the efficiency. $\dagger$ denotes methods without optimized implementation and \textcolor{red}{OOM} denotes out of memory on a single RTX 3090 GPU. Blanked entries correspond to results unable to be reported.}
\label{Tab:efficient}

\centering
\scalebox{0.92}{
\begin{tabular}{c|cccccc}
        \hline
        Method & 32$\times$32 & 64$\times$64 & 128$\times$128 & 256$\times$256 & 216$\times$312 & 384$\times$384\\
        \hline
        DPT & 117.97/0.26 & 130.08/0.57 & 429.94/1.67 & 4539.51/8.60 & 4759.83/8.99 & \textcolor{red}{OOM}\\
        LF-DET & 73.97/0.35 & 387.88/3.58 & \textcolor{red}{OOM} & \textcolor{red}{OOM} & \textcolor{red}{OOM} & \textcolor{red}{OOM}\\
        LFT$\dagger$ & 146.34/1.79 & \textcolor{red}{OOM} & \textcolor{red}{OOM} & \textcolor{red}{OOM} & \textcolor{red}{OOM} & \textcolor{red}{OOM}\\
        LFT & 132.37/0.22 & 778.57/0.69 & 8828.29/3.51 & - & - & - \\
        EPIT$\dagger$ & 53.21/0.38 & 164.94/2.51 & 1019.11/18.38 & \textcolor{red}{OOM} & \textcolor{red}{OOM} & \textcolor{red}{OOM}\\
        EPIT & 51.58/0.13 & 123.61/0.47 & 752.24/1.81 & 5518.96/7.19 & \textcolor{red}{OOM} & 17852.41/16.16\\
        \hline
        Ours & 27.81/0.18 & 105.19/0.67 & 455.30/2.63 & 1855.72/10.47 & 1935.71/10.77 & 4277.76/23.55 \\
        
        \hline

\end{tabular}
}

\end{table}

\begin{table*}[t!]
\centering
\caption{Full-resolution inference (denoted by *) quantitative results (PSNR / SSIM) on the five public benchmarks of LF super-resolution with the input angular resolution of 5$\times$5.  We mark the best, second best results in \textbf{bold} and with \underline{underline}, respectively. \textcolor{red}{OOM} denotes out of memory on a single RTX 3090 GPU.}
\label{Tab:wholeSR}
\scalebox{0.75}{
\begin{tabular}{@{}lc|cc|ccccc}
\toprule
Method & Scale & Params.(M) & Time (ms) & \multicolumn{1}{c}{HCI-new} & \multicolumn{1}{c}{HCI-old} & \multicolumn{1}{c}{EPFL} & \multicolumn{1}{c}{INRIA} & \multicolumn{1}{c}{STFGantry} \\ \midrule

LFSAV* & 2$\times$ & 1.22 & 10.57  & 37.44/.9777  & 44.32/.9942  & \underline{35.65}/.9773  & \underline{38.16}/.9850  &  38.74/.9914 \\

HLFSR* & 2$\times$ & 3.45 & 54.29  & 37.97/.9798  & \underline{45.01}/.9949  & 35.16/\underline{.9779}  & 37.05/\underline{.9856}  &  40.53/.9944 \\ 
\midrule

DPT* & 2$\times$ & 3.73 & 117.97  & 37.29/.9770  & \textcolor{red}{OOM}  & 34.48/.9758  & 36.19/.9843  &  39.38/.9928 \\ 

EPIT* & 2$\times$ & 1.42 & 51.58  & \textbf{38.25/.9810}  &  \textbf{45.12}/.9949  & \textcolor{red}{OOM}  & \textcolor{red}{OOM}  & \textbf{42.22/.9957} \\ \midrule

MLFSR* (Ours) & 2$\times$ & 1.36 & 27.81  &  \underline{38.15}/\underline{.9802}  & 44.97/\textbf{.9950} & \textbf{36.01}/\textbf{.9798}  & \textbf{38.65}/\textbf{.9864}  & \underline{41.03}/\underline{.9947}  \\ \midrule \midrule

LFSAV* & 4$\times$ & 1.54 & 36.36  & 31.47/.9219  & 37.52/.9722  & \underline{29.99}/\textbf{.9226}  & 32.17/\textbf{.9534}  &  31.41/.9506 \\

HLFSR*  & 4$\times$ & 3.48 & 59.08 & 31.46/.9229 & \underline{37.77}/\underline{.9739}  & 29.94/.9217  & \underline{32.20}/\textbf{.9534}  & 31.53/.9529  \\
\midrule

DPT* & 4$\times$ & 3.78 & 118.62  & 31.11/.9188  & 37.20/.9721  & 29.40/.9171  & 31.38/.9505  &  31.00/.9489 \\

EPIT*  & 4$\times$ & 1.47 & 52.75  & \underline{31.53}/\underline{.9232} & 37.74/.9738   & 29.90/.9200  & 32.16/\underline{.9529} & \textbf{32.26}/\underline{.9573} \\\midrule

MLFSR* (Ours) & 4$\times$ & 1.41 & 28.94 &  \textbf{31.59}/\textbf{.9237}  &   \textbf{37.92}/\textbf{.9747}   & \textbf{30.04}/\underline{.9221} & \textbf{32.21}/\textbf{.9534}  & \underline{32.21}/\textbf{.9575}  \\

\bottomrule

\end{tabular}
}
\end{table*}

\subsection{Comparison with State-of-the-Arts}
We compare our method with 13 baselines methods, including 9 CNN-based methods~\cite{kim2016accurate,lim2017enhanced,zhang2018image,zhang2019residual,yeung2018light,wang2020spatial,wang2022disentangling,cheng2022spatial,van2023light} and 4 transformer-based methods~\cite{liang2022light,wang2022detail,cong2023exploiting,liang2023learning}.

\noindent\textbf{Quantitative Results.}
The quantitative results are shown in Table~\ref{Tab:SR}.
We can notice that MLFSR outperforms CNN-based methods at both $2\times$ and $4\times$ scale.
For example, MLFSR has a 0.33dB gain compared to DistgSSR~\cite{wang2022disentangling} and HLFSR\cite{van2023light} on the EPFL testset at $2\times$ scale.
Compared to Transformer-based methods, MLFSR shows competitive performance on most testsets.
When inference with full-image input, the performance further improves on most testsets at both $2\times$ and $4\times$ scale.
Specifically, the $4\times$ results even outperform all Transformer-based methods while maintaining a low number of parameters and fast inference speed.

\noindent\textbf{Efficiency Comparison.}
We provide an in depth efficiency comparison respect to input resolution between MLFSR and current Transformer-based methods on two metrics: runtime and peak memory use.
Table~\ref{Tab:efficient} clearly shows that with the increase of input resolution, the runtime of Transformer-based methods grows significantly.
The increase of peak memory use in Table~\ref{Tab:efficient} shows a similar trend.
Almost all baseline methods failed to infer at 
384$\times$384 resolution, and without memory optimization (\texttt{torch.nn.MultiheadAttention} interface), LFT cannot even handle 5$\times$5$\times$64$\times$64 inputs.
With optimized implementation, LFT enables inference at 128$\times$128 resolution, which is slower than our method inference at 384$\times$384 resolution.
Although EPIT with optimization consumes less memory, the inference speed is still much slower compared to MLFSR.
For example, at 256$\times$256 and 384$\times384$ resolution, EPIT is about 3.0$\times$ and 4.2$\times$ slower than our method, respectively.
On the other hand, when input asymmetric resolution such as 216$\times$312, optimized implementation fails to reduce the memory consumption which largely limits its practical use.
The above results indicate that our method demonstrates significant efficiency advantages as the input resolution increases. 

\noindent\textbf{Full-Resolution Inference Results.}
We also provide the full-resolution inference results of available top-performing methods in Table~\ref{Tab:wholeSR} for a more comprehensive evaluation.
With extra information from larger resolution inputs, the performance of baseline method improves in most case.
Our method still achieves state-of-the-art performance under the full-resolution inference setting with lower parameters and latency.

\begin{figure*}[!t]
  \begin{center}
  \begin{minipage}{\linewidth}
    \begin{minipage}{0.239\linewidth}
    \centerline{\includegraphics[width=1\linewidth]{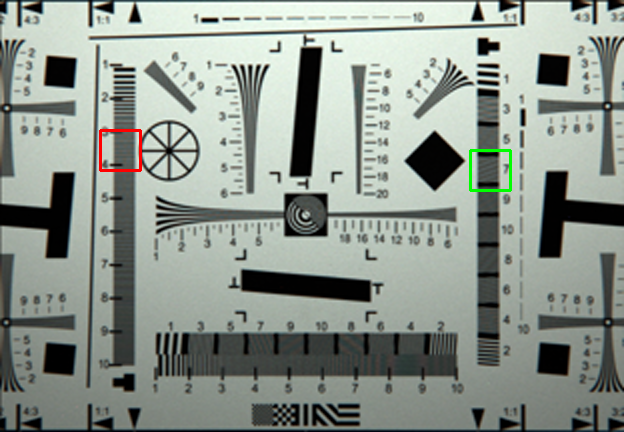}}
    \vfill
    \centerline{\includegraphics[width=0.493\linewidth]{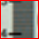} 
    \hspace{-1.8mm}
    \includegraphics[width=0.493\linewidth]{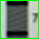}}
    \vfill \vspace{-0.15cm}
    \centerline{\scriptsize{Bicubic}}
    \end{minipage}
    \begin{minipage}{0.239\linewidth}
    \centerline{\includegraphics[width=1\linewidth]{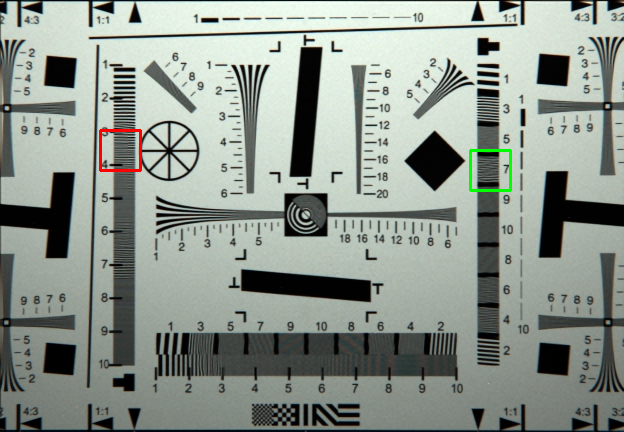}}
    \vfill
    \centerline{\includegraphics[width=0.493\linewidth]{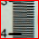} 
    \hspace{-1.8mm}
    \includegraphics[width=0.493\linewidth]{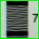}}
    \vfill \vspace{-0.15cm}
    \centerline{\scriptsize{DPT}}
    \end{minipage}
    \begin{minipage}{0.239\linewidth}
    \centerline{\includegraphics[width=1\linewidth]{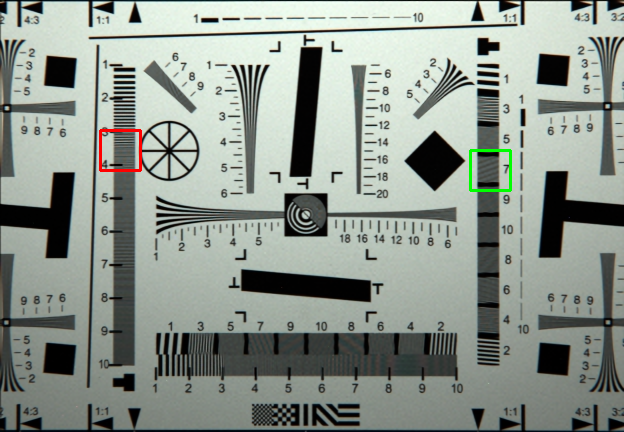}}
    \vfill
    \centerline{\includegraphics[width=0.493\linewidth]{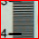} 
    \hspace{-1.8mm}
    \includegraphics[width=0.493\linewidth]{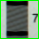}}
    \vfill \vspace{-0.15cm}
    \centerline{\scriptsize{EPIT}}
    \end{minipage}
    \begin{minipage}{0.239\linewidth}
    \centerline{\includegraphics[width=1\linewidth]{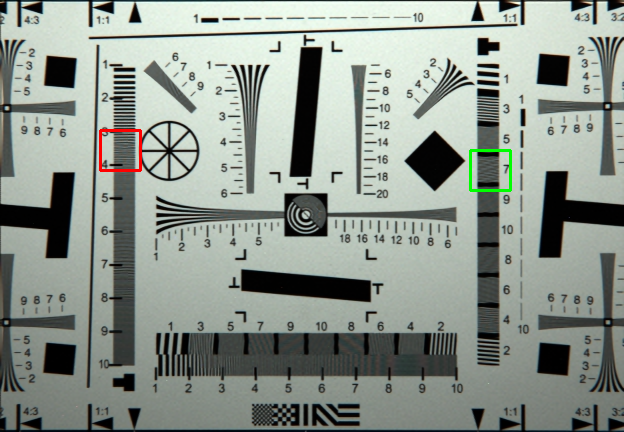}}
    \vfill
    \centerline{\includegraphics[width=0.493\linewidth]{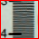} 
    \hspace{-1.8mm}
    \includegraphics[width=0.493\linewidth]{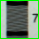}}
    \vfill \vspace{-0.15cm}
    \centerline{\scriptsize{LFT}}
    \end{minipage}
    \vfill
    \vspace{1mm}
    \begin{minipage}{0.239\linewidth}
    \centerline{\includegraphics[width=1\linewidth]{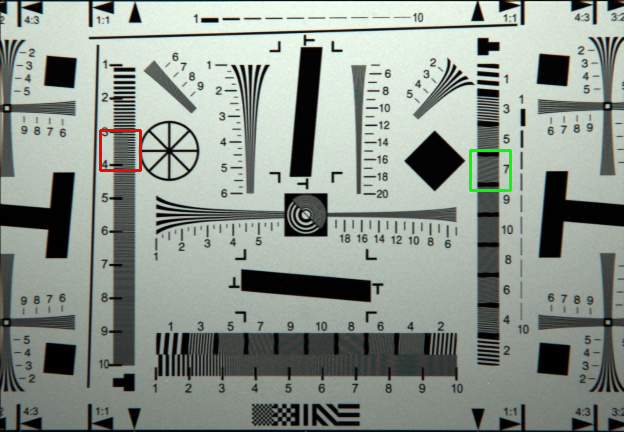}}
    \vfill
    \centerline{\includegraphics[width=0.493\linewidth]{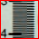} 
    \hspace{-1.8mm}
    \includegraphics[width=0.493\linewidth]{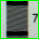}}
    \vfill \vspace{-0.15cm}
    \centerline{\scriptsize{LF-DET}}
    \end{minipage}
    \begin{minipage}{0.239\linewidth}
    \centerline{\includegraphics[width=1\linewidth]{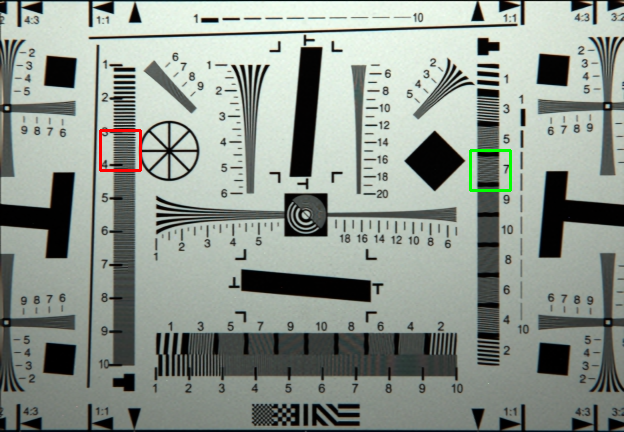}}
    \vfill
    \centerline{\includegraphics[width=0.493\linewidth]{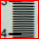} 
    \hspace{-1.8mm}
    \includegraphics[width=0.493\linewidth]{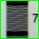}}
    \vfill \vspace{-0.15cm}
    \centerline{\scriptsize{MLFSR}}
    \end{minipage}
    \begin{minipage}{0.239\linewidth}
    \centerline{\includegraphics[width=1\linewidth]{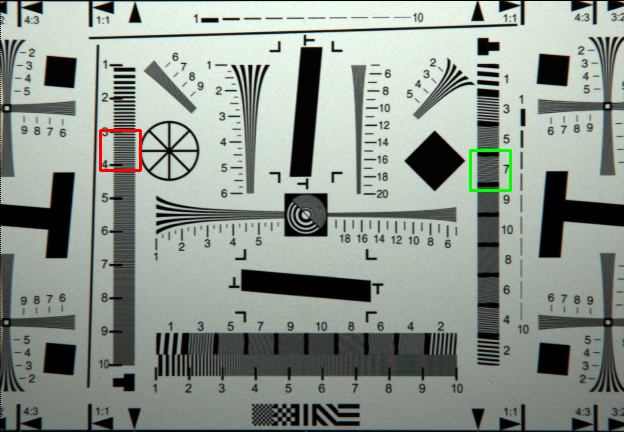}}
    \vfill
    \centerline{\includegraphics[width=0.493\linewidth]{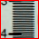} 
    \hspace{-1.8mm}
    \includegraphics[width=0.493\linewidth]{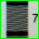}}
    \vfill \vspace{-0.15cm}
    \centerline{\scriptsize{MLFSR*}}
    \end{minipage}
    \begin{minipage}{0.239\linewidth}
    \centerline{\includegraphics[width=1\linewidth]{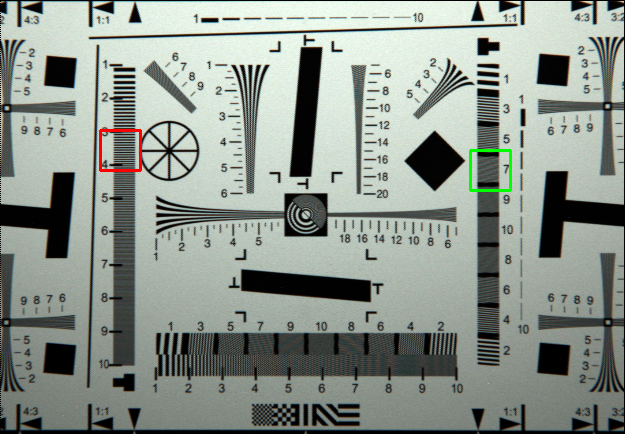}}
    \vfill
    \centerline{\includegraphics[width=0.493\linewidth]{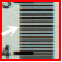} 
    \hspace{-1.8mm}
    \includegraphics[width=0.493\linewidth]{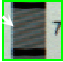}}
    \vfill \vspace{-0.15cm}
    \centerline{\scriptsize{Ground truth}}
    \end{minipage}
    \vfill
    \vspace{1mm}
      \begin{minipage}{0.239\linewidth}
    \centerline{\includegraphics[width=1\linewidth]{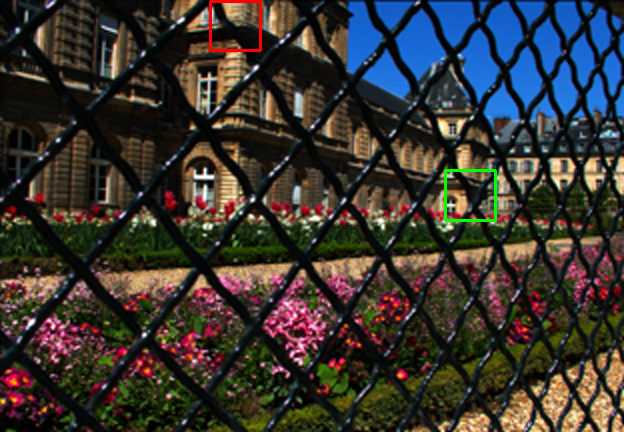}}
    \vfill
    \centerline{\includegraphics[width=0.493\linewidth]{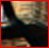} 
    \hspace{-1.8mm}
    \includegraphics[width=0.493\linewidth]{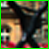}}
    \vfill \vspace{-0.15cm}
    \centerline{\scriptsize{Bicubic}}
    \end{minipage}
    \begin{minipage}{0.239\linewidth}
    \centerline{\includegraphics[width=1\linewidth]{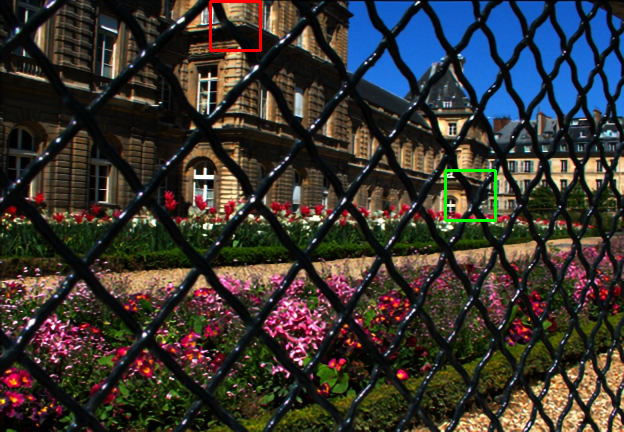}}
    \vfill
    \centerline{\includegraphics[width=0.493\linewidth]{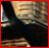} 
    \hspace{-1.8mm}
    \includegraphics[width=0.493\linewidth]{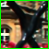}}
    \vfill \vspace{-0.15cm}
    \centerline{\scriptsize{DPT}}
    \end{minipage}
    \begin{minipage}{0.239\linewidth}
    \centerline{\includegraphics[width=1\linewidth]{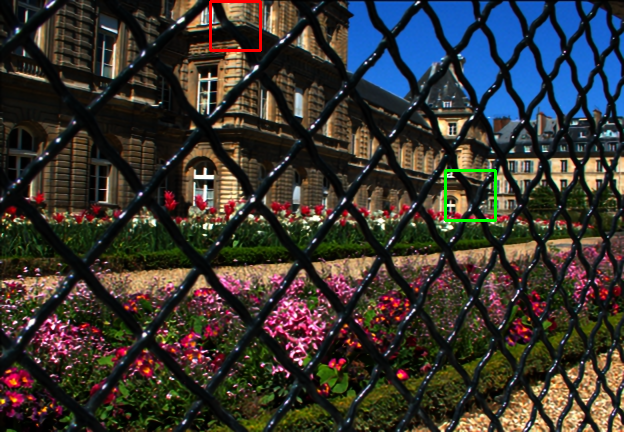}}
    \vfill
    \centerline{\includegraphics[width=0.493\linewidth]{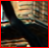} 
    \hspace{-1.8mm}
    \includegraphics[width=0.493\linewidth]{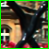}}
    \vfill \vspace{-0.15cm}
    \centerline{\scriptsize{EPIT}}
    \end{minipage}
    \begin{minipage}{0.239\linewidth}
    \centerline{\includegraphics[width=1\linewidth]{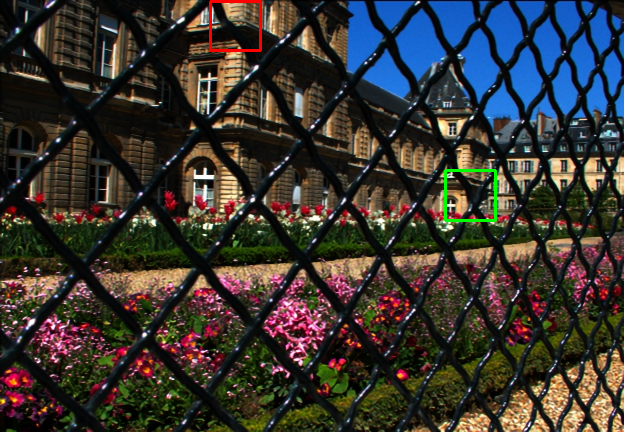}}
    \vfill
    \centerline{\includegraphics[width=0.493\linewidth]{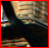} 
    \hspace{-1.8mm}
    \includegraphics[width=0.493\linewidth]{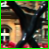}}
    \vfill \vspace{-0.15cm}
    \centerline{\scriptsize{LFT}}
    \end{minipage}
    \vfill
    \vspace{1mm}
    \begin{minipage}{0.239\linewidth}
    \centerline{\includegraphics[width=1\linewidth]{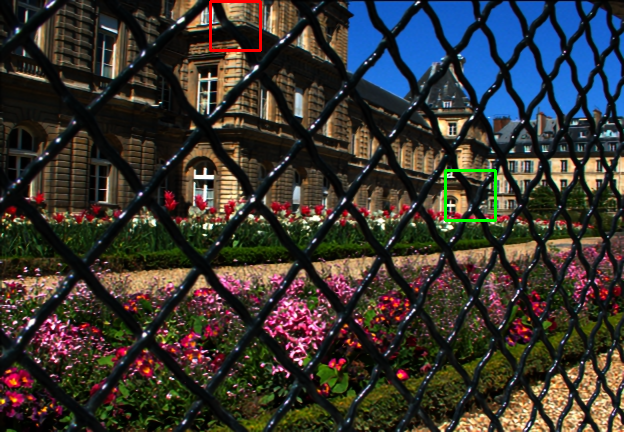}}
    \vfill
    \centerline{\includegraphics[width=0.493\linewidth]{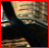} 
    \hspace{-1.8mm}
    \includegraphics[width=0.493\linewidth]{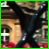}}
    \vfill \vspace{-0.15cm}
    \centerline{\scriptsize{LF-DET}}
    \end{minipage}
    \begin{minipage}{0.239\linewidth}
    \centerline{\includegraphics[width=1\linewidth]{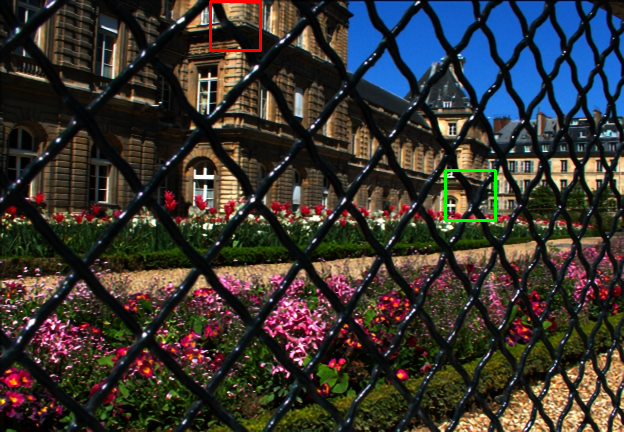}}
    \vfill
    \centerline{\includegraphics[width=0.493\linewidth]{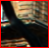} 
    \hspace{-1.8mm}
    \includegraphics[width=0.493\linewidth]{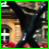}}
    \vfill \vspace{-0.15cm}
    \centerline{\scriptsize{MLFSR}}
    \end{minipage}
    \begin{minipage}{0.239\linewidth}
    \centerline{\includegraphics[width=1\linewidth]{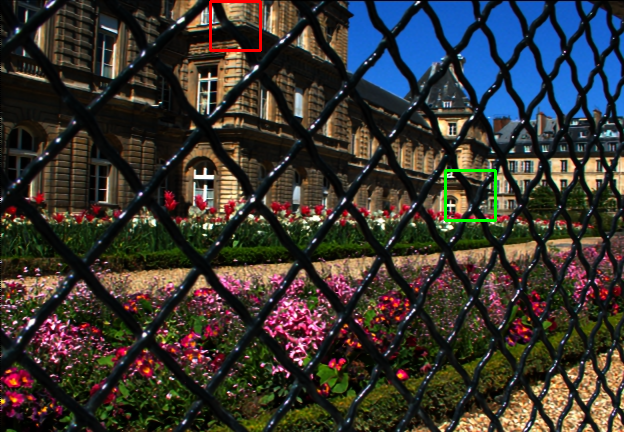}}
    \vfill
    \centerline{\includegraphics[width=0.493\linewidth]{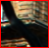} 
    \hspace{-1.8mm}
    \includegraphics[width=0.493\linewidth]{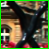}}
    \vfill \vspace{-0.15cm}
    \centerline{\scriptsize{MLFSR*}}
    \end{minipage}
    \begin{minipage}{0.239\linewidth}
    \centerline{\includegraphics[width=1\linewidth]{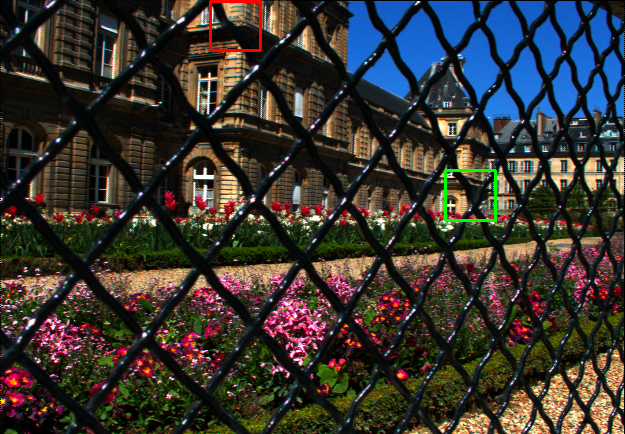}}
    \vfill
    \centerline{\includegraphics[width=0.493\linewidth]{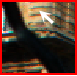} 
    \hspace{-1.8mm}
    \includegraphics[width=0.493\linewidth]{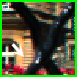}}
    \vfill \vspace{-0.15cm}
    \centerline{\scriptsize{Ground truth}}
    \end{minipage}

    \end{minipage}
  \end{center}
    \vspace{-4mm}
  \caption{
    Visual comparisons of different methods on 2$\times$ SR (view coordinates: (2, 2)).
    Please zoom in for better visualization and best viewing on screen.
  }
    \vspace{-2mm}
\label{Fig:visual1}
\end{figure*}

\begin{figure*}[!t]
  \begin{center}
  \begin{minipage}{\linewidth}
    \begin{minipage}{0.239\linewidth}
    \centerline{\includegraphics[width=1\linewidth]{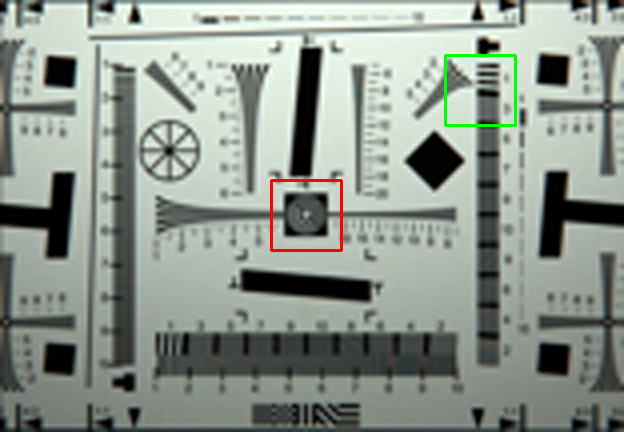}}
    \vfill
    \centerline{\includegraphics[width=0.493\linewidth]{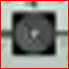} 
    \hspace{-1.8mm}
    \includegraphics[width=0.493\linewidth]{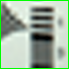}}
    \vfill \vspace{-0.15cm}
    \centerline{\scriptsize{Bicubic}}
    \end{minipage}
    \begin{minipage}{0.239\linewidth}
    \centerline{\includegraphics[width=1\linewidth]{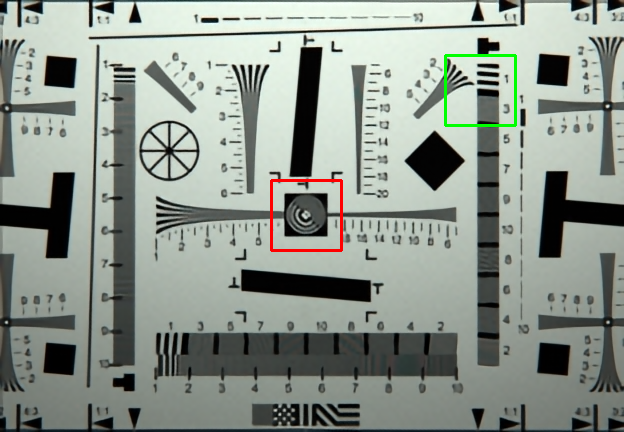}}
    \vfill
    \centerline{\includegraphics[width=0.493\linewidth]{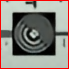} 
    \hspace{-1.8mm}
    \includegraphics[width=0.493\linewidth]{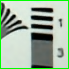}}
    \vfill \vspace{-0.15cm}
    \centerline{\scriptsize{DPT}}
    \end{minipage}
    \begin{minipage}{0.239\linewidth}
    \centerline{\includegraphics[width=1\linewidth]{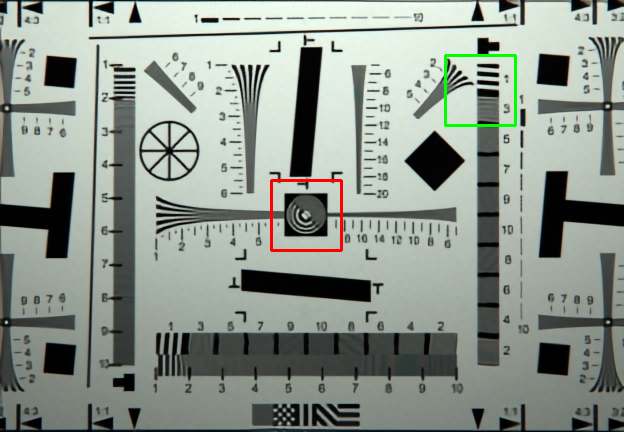}}
    \vfill
    \centerline{\includegraphics[width=0.493\linewidth]{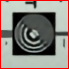} 
    \hspace{-1.8mm}
    \includegraphics[width=0.493\linewidth]{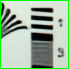}}
    \vfill \vspace{-0.15cm}
    \centerline{\scriptsize{EPIT}}
    \end{minipage}
    \begin{minipage}{0.239\linewidth}
    \centerline{\includegraphics[width=1\linewidth]{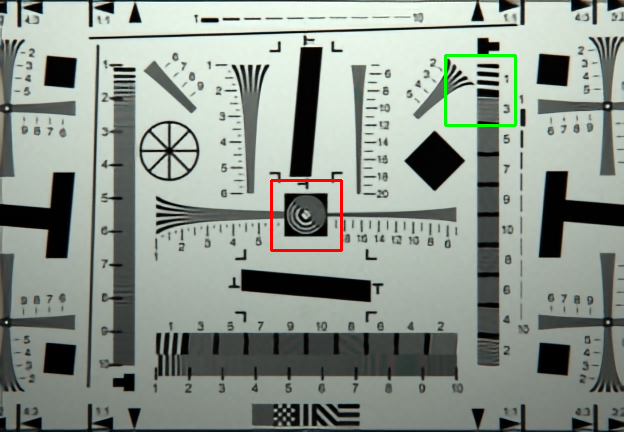}}
    \vfill
    \centerline{\includegraphics[width=0.493\linewidth]{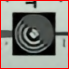} 
    \hspace{-1.8mm}
    \includegraphics[width=0.493\linewidth]{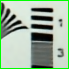}}
    \vfill \vspace{-0.15cm}
    \centerline{\scriptsize{LFT}}
    \end{minipage}
    \vfill
    \vspace{1mm}
    \begin{minipage}{0.239\linewidth}
    \centerline{\includegraphics[width=1\linewidth]{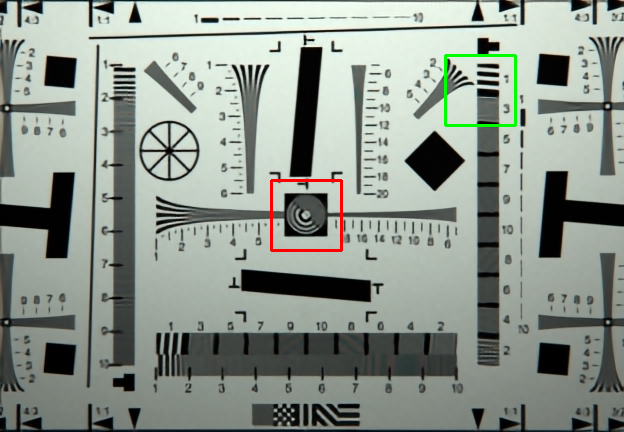}}
    \vfill
    \centerline{\includegraphics[width=0.493\linewidth]{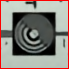} 
    \hspace{-1.8mm}
    \includegraphics[width=0.493\linewidth]{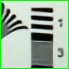}}
    \vfill \vspace{-0.15cm}
    \centerline{\scriptsize{LF-DET}}
    \end{minipage}
    \begin{minipage}{0.239\linewidth}
    \centerline{\includegraphics[width=1\linewidth]{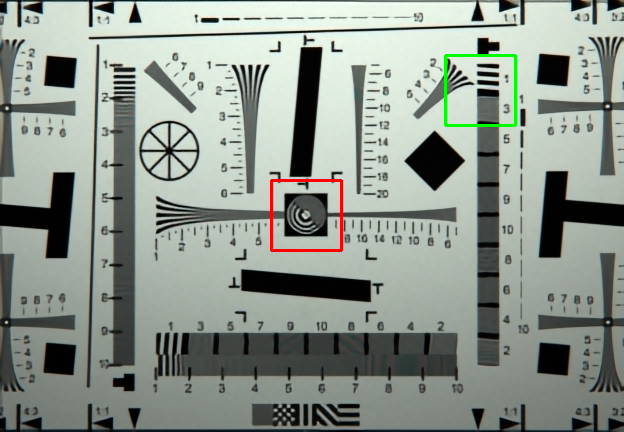}}
    \vfill
    \centerline{\includegraphics[width=0.493\linewidth]{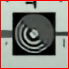} 
    \hspace{-1.8mm}
    \includegraphics[width=0.493\linewidth]{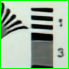}}
    \vfill \vspace{-0.15cm}
    \centerline{\scriptsize{MLFSR}}
    \end{minipage}
    \begin{minipage}{0.239\linewidth}
    \centerline{\includegraphics[width=1\linewidth]{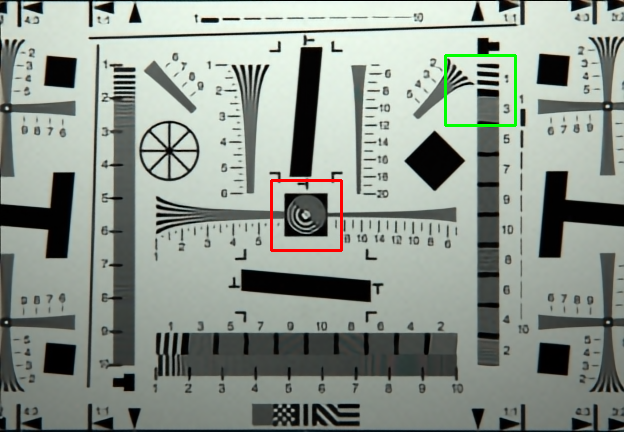}}
    \vfill
    \centerline{\includegraphics[width=0.493\linewidth]{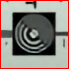} 
    \hspace{-1.8mm}
    \includegraphics[width=0.493\linewidth]{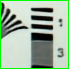}}
    \vfill \vspace{-0.15cm}
    \centerline{\scriptsize{MLFSR*}}
    \end{minipage}
    \begin{minipage}{0.239\linewidth}
    \centerline{\includegraphics[width=1\linewidth]{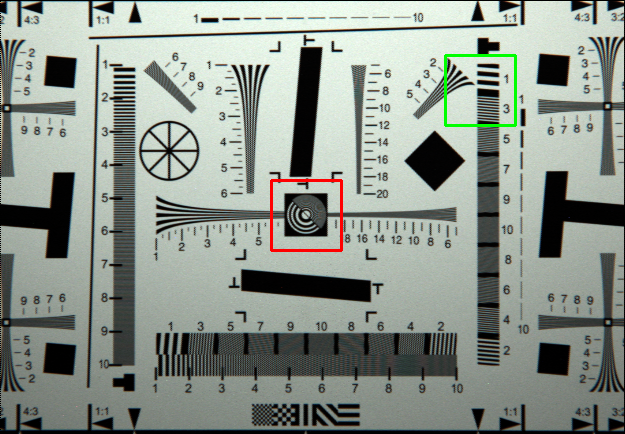}}
    \vfill
    \centerline{\includegraphics[width=0.493\linewidth]{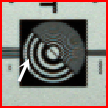} 
    \hspace{-1.8mm}
    \includegraphics[width=0.493\linewidth]{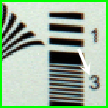}}
    \vfill \vspace{-0.15cm}
    \centerline{\scriptsize{Ground truth}}
    \end{minipage}
    \vfill
    \vspace{1mm}
    \begin{minipage}{0.239\linewidth}
    \centerline{\includegraphics[width=1\linewidth]{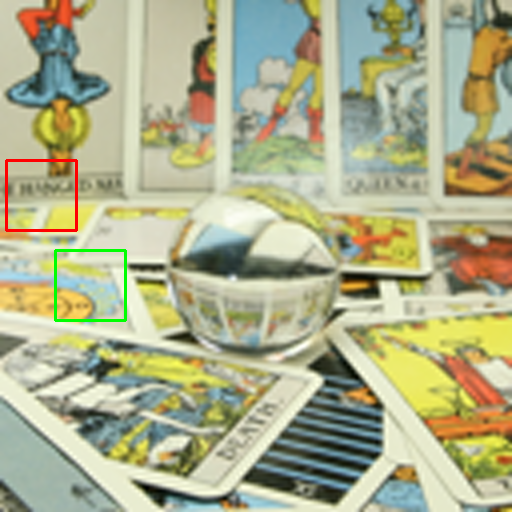}}
    \vfill
    \centerline{\includegraphics[width=0.493\linewidth]{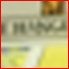} 
    \hspace{-1.8mm}
    \includegraphics[width=0.493\linewidth]{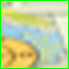}}
    \vfill \vspace{-0.15cm}
    \centerline{\scriptsize{Bicubic}}
    \end{minipage}
    \begin{minipage}{0.239\linewidth}
    \centerline{\includegraphics[width=1\linewidth]{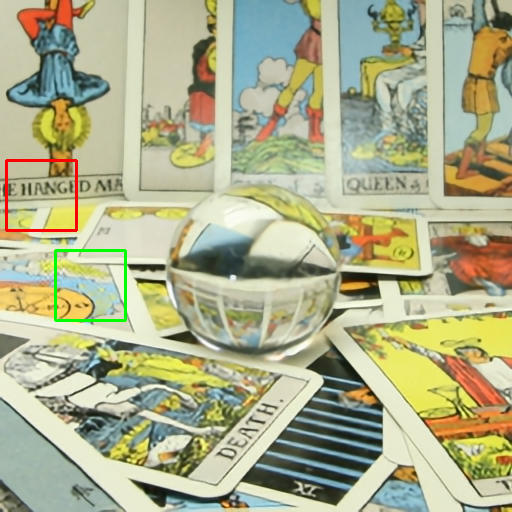}}
    \vfill
    \centerline{\includegraphics[width=0.493\linewidth]{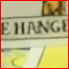} 
    \hspace{-1.8mm}
    \includegraphics[width=0.493\linewidth]{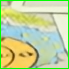}}
    \vfill \vspace{-0.15cm}
    \centerline{\scriptsize{DPT}}
    \end{minipage}
    \begin{minipage}{0.239\linewidth}
    \centerline{\includegraphics[width=1\linewidth]{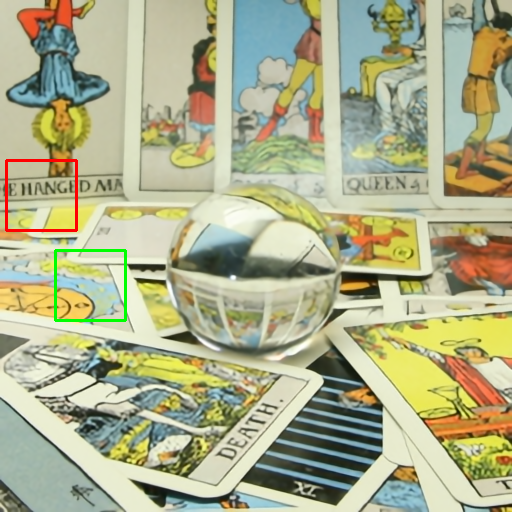}}
    \vfill
    \centerline{\includegraphics[width=0.493\linewidth]{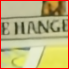} 
    \hspace{-1.8mm}
    \includegraphics[width=0.493\linewidth]{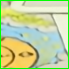}}
    \vfill \vspace{-0.15cm}
    \centerline{\scriptsize{EPIT}}
    \end{minipage}
    \begin{minipage}{0.239\linewidth}
    \centerline{\includegraphics[width=1\linewidth]{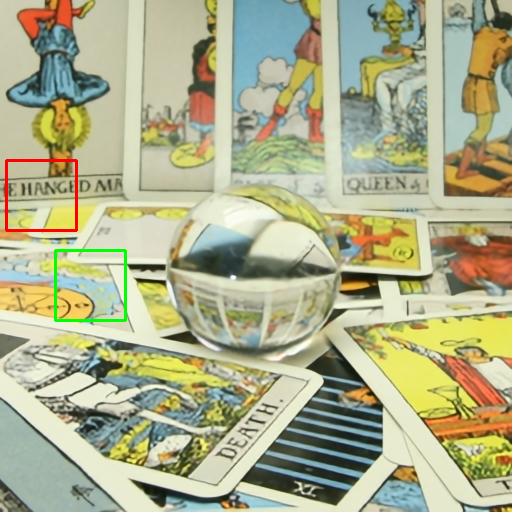}}
    \vfill
    \centerline{\includegraphics[width=0.493\linewidth]{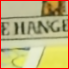} 
    \hspace{-1.8mm}
    \includegraphics[width=0.493\linewidth]{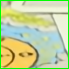}}
    \vfill \vspace{-0.15cm}
    \centerline{\scriptsize{LFT}}
    \end{minipage}
    \vfill
    \vspace{1mm}
    \begin{minipage}{0.239\linewidth}
    \centerline{\includegraphics[width=1\linewidth]{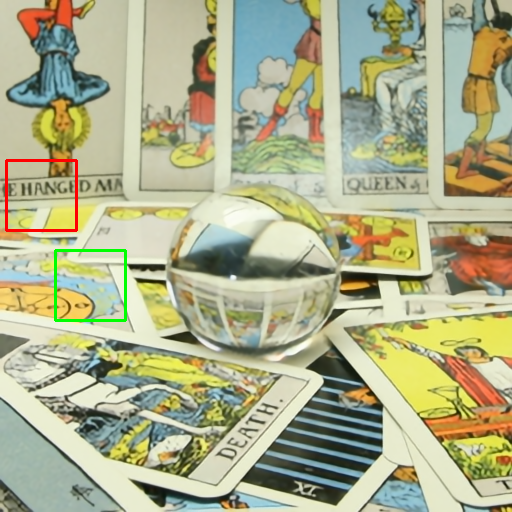}}
    \vfill
    \centerline{\includegraphics[width=0.493\linewidth]{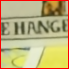} 
    \hspace{-1.8mm}
    \includegraphics[width=0.493\linewidth]{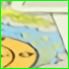}}
    \vfill \vspace{-0.15cm}
    \centerline{\scriptsize{LF-DET}}
    \end{minipage}
    \begin{minipage}{0.239\linewidth}
    \centerline{\includegraphics[width=1\linewidth]{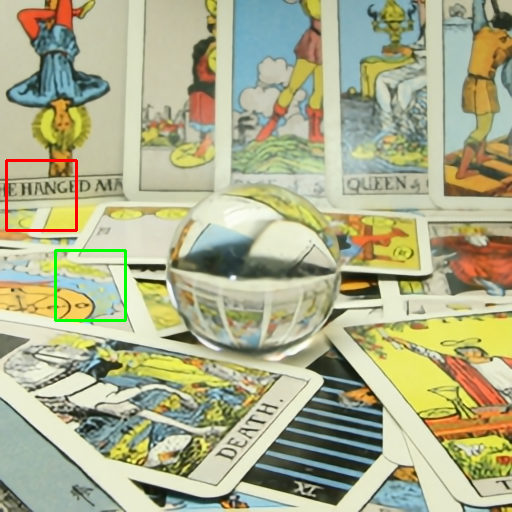}}
    \vfill
    \centerline{\includegraphics[width=0.493\linewidth]{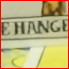} 
    \hspace{-1.8mm}
    \includegraphics[width=0.493\linewidth]{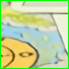}}
    \vfill \vspace{-0.15cm}
    \centerline{\scriptsize{MLFSR}}
    \end{minipage}
    \begin{minipage}{0.239\linewidth}
    \centerline{\includegraphics[width=1\linewidth]{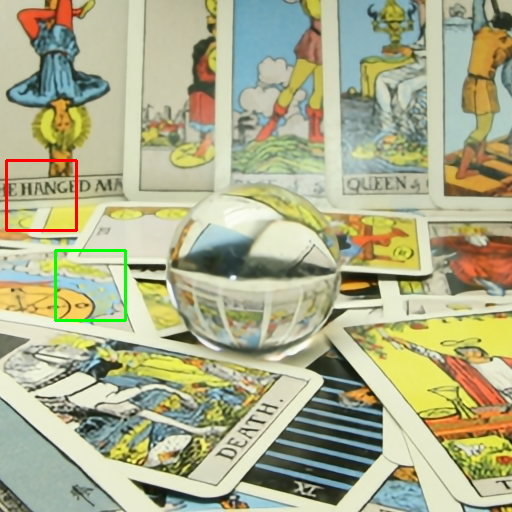}}
    \vfill
    \centerline{\includegraphics[width=0.493\linewidth]{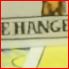} 
    \hspace{-1.8mm}
    \includegraphics[width=0.493\linewidth]{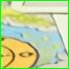}}
    \vfill \vspace{-0.15cm}
    \centerline{\scriptsize{MLFSR*}}
    \end{minipage}
    \begin{minipage}{0.239\linewidth}
    \centerline{\includegraphics[width=1\linewidth]{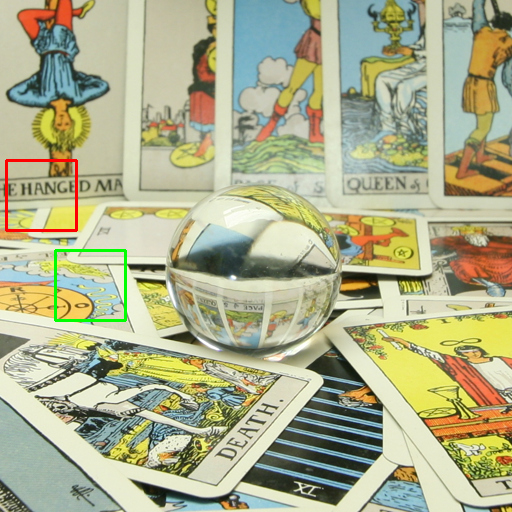}}
    \vfill
    \centerline{\includegraphics[width=0.493\linewidth]{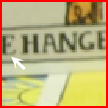} 
    \hspace{-1.8mm}
    \includegraphics[width=0.493\linewidth]{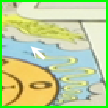}}
    \vfill \vspace{-0.15cm}
    \centerline{\scriptsize{Ground truth}}
    \end{minipage}
    
    \end{minipage}
  \end{center}
  \vspace{-4mm}
  \caption{
    Visual comparisons of different methods on 4$\times$ SR (view coordinates: (2, 2)).
    Please zoom in for better visualization and best viewing on screen.
  }
  \vspace{-2mm}
  \label{Fig:visual2}
\end{figure*}

\noindent\textbf{Qualitative Results.} 
We provide visual results on 2$\times$ and 4$\times$ scale in Fig.~\ref{Fig:visual1} and  Fig.~\ref{Fig:visual2}, respectively.
For example, in the ISO scene in Fig.~\ref{Fig:visual1}, our results manages to recover the horizontal line.
Note that the result of DPT and LF-DET contains vertical line artifacts due to the patch inference scheme, which shows the superiority of whole-image inference.
The details on the background wall in the below scene are also recovered by our method.
In Fig.~\ref{Fig:visual2}, our method successfully produces clear number 3 in the ISO scene while other baselines tend to generate blurry results and artifacts.
The letters in the tarot card scene generated by our method have less distorted edges and are more close to the ground truth.

\subsection{Ablation Studies}
\label{Sec:ablation}
In this section, we perform ablation studies on 4$\times$ SR to validate the effectiveness of our designs. 
Specifically, we consider the core component of MLFSR, \textit{i.e.,} SA-Mamba, EPI-Mamba, SAM, and the T2M distillation loss
The quantitative results are shown in Table~\ref{Tab:ab1}.
By removing either SA-Mamba or EPI-Mamba in MLFSR, the performance dropped by 0.14dB and 0.18dB on the average PSNR respectively, showing the importance of complementary information between subspaces.
On the other hand, without the locality introduced by SAM, the performance also dropped by 0.11dB, indicating the important role of local information in the SR task.
Finally, the T2M loss further boosts the performance by 0.06dB.
The above results validate the effectiveness of our core designs.

\begin{table}[!t]
\caption{Ablations on the macro design. The best results are marked in \textbf{bold}. All methods are under full-resolution inference scheme.}
\label{Tab:ab1}

\centering
\scalebox{0.95}{
\begin{tabular}{cccc|c}
        \hline
        EPI-Mamba & SA-Mamba & SAM & T2M Loss & Avg. PSNR/SSIM \\
        \hline
        $\usym{2717}$ & $\usym{2713}$ & $\usym{2713}$ & $\usym{2713}$ & 32.65/.9457 \\
        $\usym{2713}$ & $\usym{2717}$ & $\usym{2713}$ & $\usym{2713}$ & 32.60/.9454 \\
        $\usym{2713}$ & $\usym{2713}$ & $\usym{2717}$ & $\usym{2713}$ & 32.68/.9456 \\
        $\usym{2713}$ & $\usym{2713}$ & $\usym{2713}$ & $\usym{2717}$ & 32.73/.9460 \\
        \hline
        $\usym{2713}$ & $\usym{2713}$ & $\usym{2713}$ & $\usym{2713}$ & 32.79/.9463 \\
        \hline

\end{tabular}
}

\end{table}

\section{Conclusion}
In this paper, we propose a Mamba-based LFSR method named MLFSR to ease the high memory consumption and latency brought by Transformer-based methods.
Specifically, we utilize the inherent structure redundancy existing in LFs and propose an efficient subspace scanning method.
Based on it, we design a Mamba-based Global Interaction module to model global spatial-angular correlations. 
Furthermore, a Spatial-Angular Modulator is proposed to complement local information.
The overall performance is further boosted by a Transformer-to-Mamba distillation loss.
Extensive experiments show that MLFSR has a clear advantage in efficiency while achieving state-of-art performance. 
With much less memory use, MLFSR facilitates further performance improvement with full-resolution inference.


\bibliographystyle{splncs04}
\bibliography{main}
\end{document}